\documentclass[11pt]{article}

\pdfoutput=1

\usepackage{mdframed} 

\usepackage{esvect}

\usepackage{cite}
\usepackage{hyperref}
\usepackage{soul}

\usepackage[dvipsnames]{xcolor}
\usepackage[normalem]{ulem}

\usepackage{lineno}

\usepackage{amsmath}
\usepackage{bbold}
\usepackage{amssymb,amsfonts,amsthm}
\usepackage{epsfig,graphicx}
\usepackage{url} 
\usepackage{bm} 
\usepackage{slashed}
\usepackage{cite}

\usepackage{color}

\usepackage{fancyhdr}
\usepackage{textcomp}

\usepackage{float}
\usepackage{caption}
\usepackage{subcaption}
\usepackage{enumitem}
\usepackage{multirow}

\def \GeV{{\mathrm{GeV}}}
\def \TeV{{\mathrm{TeV}}}

\DeclareMathAlphabet{\pazocal}{OMS}{zplm}{m}{n}

\numberwithin{equation}{section}

\definecolor{mypurple}{HTML}{6A0888}

\begin{document}

\begin{flushright}
IFT-UAM/CSIC-20-25
\end{flushright}

\begin{center}

  {\bf {\LARGE Improving  $t \bar{t}$ reconstruction  in the dilepton channel \\[1mm] at future lepton colliders}}
  
\renewcommand*{\thefootnote}{\fnsymbol{footnote}}
\setcounter{footnote}{0}

  \vspace{0.5cm}
  {\large
    P.~Mart\'\i n-Ramiro $^{a}$ 
                                           \footnote{ORCID \href{https://orcid.org/0000-0001-5858-5783}{0000-0001-5858-5783}}, and
    J.~M. Moreno $^{a}$ 
                                          \footnote{ORCID  \href{https://orcid.org/0000-0002-2941-0690}{0000-0002-2941-0690}}
  }
  \\[0.2cm] 

  {\footnotesize{
$^a$ Instituto de F\'\i sica Te\'orica, IFT-UAM/CSIC \\ Universidad Aut\'onoma de Madrid, Madrid 28049, Spain
        }
    }

\vspace*{0.7cm}

\begin{abstract}
  A future lepton collider, such as the proposed CLIC or ILC, would allow to study top quark properties with unprecedented precision. In this paper, we present a method to reconstruct the $t \bar{t}$ decay in the dilepton channel at future $e^+e^-$ colliders. We derive a simple, closed analytical expression for the neutrino four-momenta as a function of the $W$ boson mass and develop a maximization procedure to find the optimal solution for the reconstruction of the full $t \bar{t}$ event. We show that our method is able to reconstruct neutrino four-momenta with an error of less than $2 \, \%$ in $60 \, \%$ of the times. Finally, we test the performance of this reconstruction method in the calculation of the helicity fractions of the $W$ boson. A precise measurement of these observables could be used to probe new physics effects in the $Wtb$ vertex. We find that, from  a large $t \bar{t}$ sample, our reconstruction method allows to calculate these observables with an accuracy better than $1 \, \%$.
\end{abstract}
\end{center}

\newpage

\section{Introduction}
\label{sec:Intro}

The top quark, being the heaviest fermion, plays a special role in the Standard Model (SM). Loops involving the top quark are crucial in electroweak physics, providing sizeable contributions to electroweak precision observables. They also induce the dominant Higgs production mechanism at hadron colliders and shape the clean Higgs diphoton decay mode. The next generation of $e^{+}e^{-}$ colliders will offer a great opportunity to study top quark properties with unprecedented precision. In particular, the proposed Compact Linear Collider (CLIC) \cite{Aicheler:2012bya} and International Linear Collider (ILC) \cite{Behnke:2013xla} would allow to make high-precision measurements of the top quark mass, its decay products and polarization, and potentially observe signals of new physics (NP) that may couple to top quarks \cite{Baer:2013cma, Charles:2018vfv, Abramowicz:2018rjq,  Fujii:2019zll, Bambade:2019fyw}.

In order to perform such measurements, one needs to reconstruct the top quark from its decay products accurately and efficiently. The $e^+e^- \rightarrow t \bar{t}$ production process, with the top quarks decaying to a $W$ boson and a bottom quark, has a significant cross section above the $t \bar{t}$ production threshold \cite{Fujii:2019zll}.  Although the dilepton channel, where the two $W$ bosons decay to leptons  is cleaner than the hadronic channel, the neutrino reconstruction can be challenging. In hadronic colliders, including the Large Hadron Collider (LHC) and Tevatron, several kinematic reconstruction algorithms such as the Neutrino Weighting \cite{Varnes:1997tc} and Matrix Element Weighting \cite{Dalitz:1991wa} methods have been used with success by the different collaborations (ATLAS \cite{Aaboud:2016syx}, CDF \cite{Aaltonen:2015hta} , CMS \cite{Khachatryan:2015oqa, Sirunyan:2018ucr}, D0 \cite{ D0:2015dxa, D0:2016ull}).

In this work, we develop a new approach to perform the analytical reconstruction of top quark pairs decaying in the dilepton channel at CLIC and ILC. 
Imposing energy-momentum conservation is not enough to fix the two neutrino three-momenta, so an educated ansatz is needed to set the two remaining parameters.
First, we solve these conservation equations assuming that the $W$ bosons produced in the decay of the top quarks are on-shell and their masses are fixed to the pole value, providing a simple analytical solution (Method I). After that, we introduce a weight function proportional to the top quark and $W$ boson propagators (and therefore the amplitude for the process) to improve the reconstruction, allowing for the reconstructed top and $W$ masses to vary (Method II). We find that both methods provide highly accurate results for the reconstruction. In particular, Method I has the advantage of being computationally very fast, while Method II provides an extremely precise reconstruction of the event, at the cost of being computationally slower.

This paper is organised as follows. In Section~\ref{sec:MethodI}, we explicitly solve the system of kinematical equations analytically for the studied process and present Method I, with the goal of reconstructing the full $t\bar{t}$ event. After that, we develop an improved version of the reconstruction, Method II, and analyse the performance of both approaches in Section~\ref{sec:MethodII}. Later, as an application, we show how these techniques can be used for measuring the $W$ boson polarization, and therefore, to probe new physics involving the $Wtb$ coupling. Finally, we present the conclusions of this work in Section~\ref{sec:conclusions}.

\section{Method I: On-shell reconstruction}
\label{sec:MethodI}

In this section we solve the kinematic equations of the $ t {\bar t} $ decay in the dilepton channel, assuming that the $W$ bosons that mediate the decay are on-shell. We derive a simple analytical expression for the neutrino four-momentum and, therefore, reconstruct the full event. Note that the reconstruction in $e^+e^-$ colliders is, a priori, more accurate than in hadronic colliders\footnote{An analytical solution to the kinematic equations was provided in \cite{Sonnenschein:2006ud}. For a previous, geometrical method see \cite{Dalitz:1992np,Dalitz:1992bx}.}, where the energy and momentum of the initial state are unknown and therefore extra assumptions (such as on-shellness  of $ t, \bar t $ quarks) must be made.

We will denote by $P = (E, \vec{p} \,)$ the different four-momenta  and by  $p^{i}$ the individual components of $\vec{p}$,  the spatial three-momentum. To begin with, we can write the energy-momentum conservation equation as
\begin{equation}\label{eqn:conservation}
  P_{j_{1}} + P_{j_{2}} + P_{\nu} + P_{\bar \nu} + P_{l} + P_{\bar l} = P_{0} \, ,
\end{equation}
where  $j_{1, 2}$ stand for the two jets of the event and $P_{0} = (\sqrt{s}, 0, 0, 0 )$. The neutrino and anti-neutrino on-shell conditions are given by the following equations
\begin{eqnarray}\label{eqn:nu}
 P_{\nu}^{2} &=& 0 \, , \nonumber \\
 P_{\bar \nu}^{2}  = ( P_{0} - P_{\nu} - P_{l} - P_{\bar l} - P_{j_{1}} - P_{j_{2}})^{2}  &=& 0 \, .
\end{eqnarray}
Two more equations are needed  to fully reconstruct $P_{\nu}$. If we assume that the $W^{\pm}$ bosons are on-shell, their pole masses can be used as a constraint
\begin{eqnarray}\label{eqn:WW}
( P_{\nu} +  P_{\bar l})^{2}  &=&   m_{W^{+}}^{2}  \, , \nonumber \\
  (P_{0}  -  P_{\nu} - P_{\bar l} -  P_{j_{1}}  -  P_{j_{2}})^{2}  &=&   m_{W^{-}}^{2} \, .
\end{eqnarray}
Thus, we have a system of four {\em quadratic} equations, \eqref{eqn:nu}, \eqref{eqn:WW}; and four variables, $(E_{\nu}, \vec{p}_{\nu})$. Now we can greatly simplify this system of equations and derive a set of {\em linear} equations in $p^{i}_{\nu}$ by considering the equivalent system of equations
\begin{eqnarray}\label{eqn:equivalent}
P_{\nu}^{2} - P_{W^{+}}^{2} &=& - \, m_{W^{+}}^{2}  \, , \nonumber \\
P_{\bar \nu}^{2} - P_{W^{-}}^{2} &=& - \, m_{W^{-}}^{2} \, , \nonumber \\
P_{W^{+}}^{2} - P_{W^{-}}^{2} &=& m_{W^{+}}^{2} - m_{W^{-}}^{2} \, , \nonumber \\
P_{\nu}^{2} &=& 0 \, .
\end{eqnarray}
Thus, for the first 3 equations in system \eqref{eqn:equivalent} we obtain the matrix equation
\begin{equation}\label{matrixeq}
2 K_{ij} p^{j}_{\nu} = E_{\nu} a^{j}  + b^{j} \, ,
\end{equation}
with
\begin{equation}
K = 
\left( {\begin{array}{ccc}
    p_{\bar l}^{1} &  p_{\bar l}^{2} & p_{\bar l}^{3} \\
    p_{l}^{1} &  p_{l}^{2} & p_{l}^{3} \\
    p_{j_{1}} ^{1} +  p_{j_{2}} ^{1} &  p_{j_{1}}^{2} +  p_{j_{2}} ^{2}&  p_{j_{1}} ^{3} +  p_{j_{2}}^{3}  \\
  \end{array} } \right) \, ,
\end{equation}

\begin{equation}
a  = 2
\left( {\begin{array}{c}
    E_{\bar l} \\
    E_{l} \\
    E_{j_{1}}^{1} +  E_{j_{2}} ^{1}-  \sqrt{s} 
  \end{array} } 
  \right),
      \; 
b = 
\left( {\begin{array}{c}
    P_{\bar l} \cdot P_{\bar l} - m_{W^{+}}^{2} \\
    P_{l} \cdot ( P_{l} +2 (P_{\bar l} +  P_{j_{1}} +  P_{j_{2}}  - P_{0} ))   + m_{W^{-}}^{2} \\
   ( P_{j_{1}} +  P_{j_{2}}  -  P_{0}) \cdot (  2 P_{\bar l} + P_{j_{1}} +  P_{j_{2}} - P_{0}) + m_{W^{+}}^{2} - m_{W^{-}}^{2} \\
  \end{array} } 
  \right) \, .
  \end{equation}
In this way, solving eq. \eqref{matrixeq} for the individual spatial components of the neutrino three-momentum we find
\begin{equation}\label{eq:alphabeta}
  p^{i}_{\nu}  = E_{\nu}  \alpha^{i}  + \beta^{i} \, ,
\end{equation}
with  $\alpha^{i}  = \frac{1}{2} K^{-1}_{ij} a ^{j}$ , $\beta^{i}  =\frac{1}{2}  K^{-1}_{ij}  b^{j}$. Note that only the components $\beta^{i}$ depend on $m_{W^{\pm}}$. Using eq. \eqref{eq:alphabeta} we can rewrite the neutrino four-momentum as
\begin{equation}
   P_{\nu}  =  ( E_{\nu} ,  p^{i}_{\nu} ) = (E_{\nu} ,   E_{\nu}  \alpha^{i}  + \beta^{i}  ) \, ,
\end{equation}
and use the last equation of system \eqref{eqn:equivalent} to derive the following quadratic equation in $E_{\nu}$
\begin{equation}
E_{\nu}^{2} (1 - \vec{\alpha}^{\, 2}) - 2 E_{\nu} (\vec{\alpha}\cdot\vec{\beta}) - \vec{\beta}^{\,2}  = 0 \, ,
\end{equation} 
which yields two solutions
\begin{equation}
 E_{\nu}^{\pm} =  \frac{\vec{\alpha}\cdot\vec{\beta} \pm \sqrt{(\vec{\alpha}\cdot\vec{\beta})^{2} + (1 - \vec{\alpha}^{\, 2}) \vec{\beta}^{\,2}}    }
                      { 1 - \vec{\alpha}^{\, 2}}  \, .                
\label{eqn:Enu}
\end{equation}
Finally, using eq. \eqref{eq:alphabeta} we find that the neutrino momentum is given by
\begin{equation}
P_{\nu}^{\pm} = ( E_{\nu}^{\pm},\; E_{\nu}^{\pm} \alpha^{i}  + \beta^{i} ) \, ,
\label{eqn:Pnu}
\end{equation}
with $E_{\nu}$ calculated in eq. \eqref{eqn:Enu}.

In the reconstruction of the $t, \bar{t}$  four-momenta there are two combinatorial ambiguities, coming from the two possible jet assignments for the $b$ and $\bar{b}$ quarks and the two signs in eq. \eqref{eqn:Enu}, that lead to four possible reconstructions. As a consequence, one must choose a criteria to define the best reconstruction. Following the steps of \cite{Casler:2019xqn}, 
where a numerical reconstruction algorithm is presented,
we evaluate  the $\chi^{2}$ function
\begin{equation}
\chi^{2} =  \displaystyle \frac{(m_T- m_t)^{2}}{\Gamma_{t}^{2}} + \frac{(m_{\bar T}- m_t)^{2}}{\Gamma_{t}^{2}} \, ,
\label{eq:chi2}
\end{equation}
for the four cases and take the reconstruction that provides the lowest $\chi^{2}$ value. In the above definition, we denote the reconstructed top masses by $m_{T, \bar T}$ and the pole masses by $m_t$.

In order to check the quality and efficiency of this method, we performed a Monte Carlo simulation to generate one million samples of $t\bar{t}$ pairs at leading order using \texttt{MadGraph5\_aMC@NLO~2.6.3}\cite{Alwall:2014hca}. The samples were generated at partonic level using the decay-chain syntax, at a center of mass energy of $1 \; \TeV$. We remark that the goal of this study is to prove the validity and efficiency of this reconstruction method. Therefore, the analysis of any potential effects from the parton shower and detector simulation are beyond the scope of this work and left for future studies. Since the reconstruction algorithm is based on simple analytical functions it is fast and easy to implement. For reference, the analysis of the full sample of one million events only took $20$ minutes on a laptop with an Intel i7-7700HQ CPU and $8 \; \text{GB}$ of RAM.

\begin{figure}[ht!]
   \begin{center}
        \includegraphics[scale=0.475]{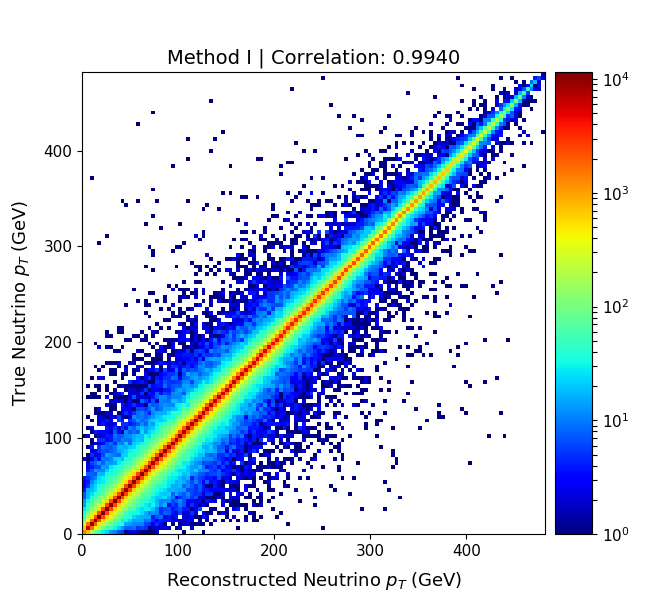}
        \includegraphics[scale=0.475]{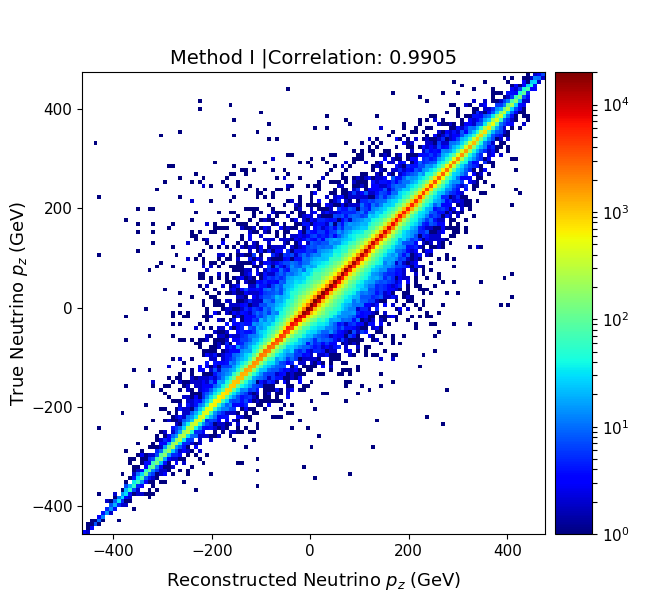}
        \includegraphics[scale=0.475]{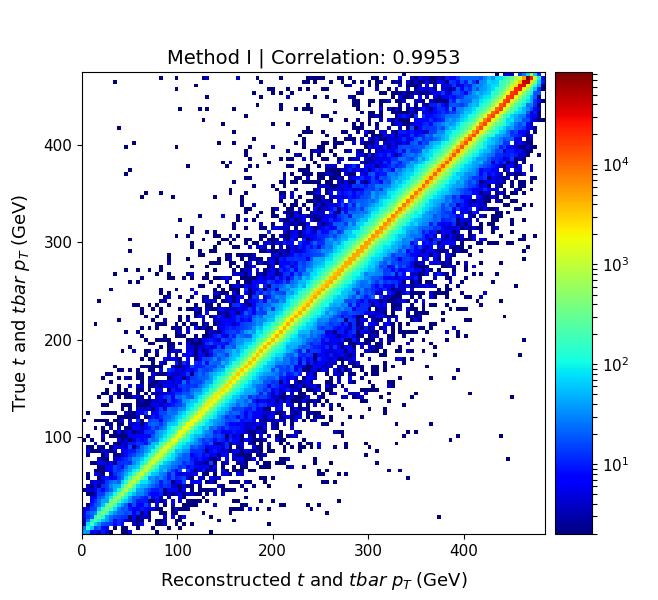}
        \includegraphics[scale=0.475]{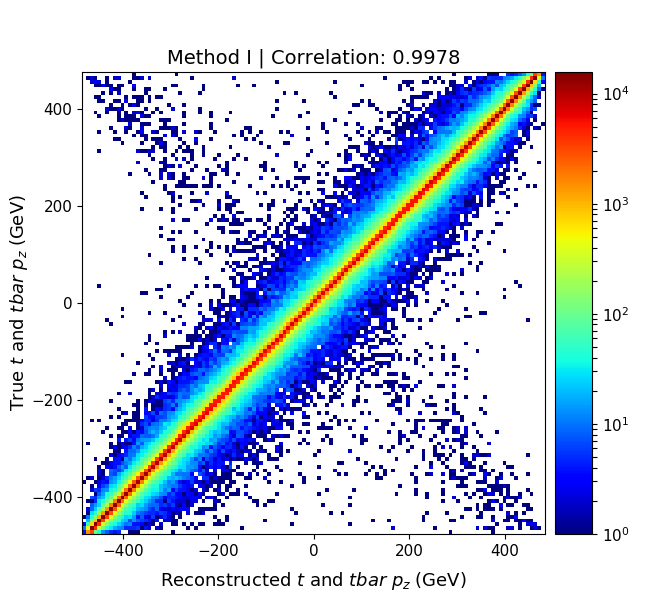}
       \end{center}
   \caption{Transverse (left) and longitudinal (right) components of the simulated neutrino (top panels) and top quarks (lower panels) spatial three-momenta as a function of the reconstructed momenta for a sample of $10^{6}$ events using Method I. The correlation between the simulated and reconstructed data is shown in every figure title.}
    \label{fig:p_standard}
\end{figure}

The outcome of the reconstruction for the neutrino and top quark momenta is presented in Fig.~\ref{fig:p_standard}. In order enable direct comparison to previous works, we show our results in the same format as \cite{Casler:2019xqn} to show the neutrino momentum reconstruction in the transverse $(p_T)$ and parallel $(p_z)$ components in the upper panels. Following the steps of \cite{Casler:2019xqn},
we evaluate the quality of the reconstruction by measuring the correlation between the simulated and reconstructed neutrino four-momenta. As can be seen in this figure, the correlation is above $0.99$ in both cases. On the other hand, the results for the top quark reconstruction are shown in the lower panels. 
The events that lie in the diagonal band that runs from the upper left to the lower right of the figure, where the reconstructed $p_{z}$ seems to have the wrong sign, correspond to an incorrect identification of the $b$ and $\bar{b}$ jets \cite{Casler:2019xqn}. However, using our analytical reconstruction method the ratio of events that are misidentified is reduced to only $0.006\%$. Finally, note that in this case the correlation between both sets of data is $\sim 0.998$, even higher than for the neutrino case. The reason for this is that the magnitude of the top quark momentum is larger than the neutrino momentum, leading to a smaller relative error. In the next section we will introduce another complementary, but more intuitive and accurate way to measure the quality of the event reconstruction.
The reconstruction of the neutrino momentum leads to (unphysical) imaginary solutions in $11\%$ of the cases, as found in \cite{Casler:2019xqn}. In these events, at least one of the $W$ bosons has an invariant mass larger than the one fixed in the reconstruction.

In order to further evaluate the performance of this method, it is interesting to compare our results with those of previous works. In particular, a significant improvement is observed by looking at Figs. 5-7 from ref.~\cite{Casler:2019xqn}. This is clearly confirmed by comparing the correlation between the simulated and reconstructed $(p_T, p_z)$ values in both analysis, which are shown in Table~\ref{tab:correlations}. Since the error in the correlation coefficient is 
${\cal O }(\epsilon^2)$, where $\epsilon$ is the relative error in the reconstruction, our method improves the reconstruction error by a factor $2-3$. The reason is that in our calculations we end up with a quadratic equations, while ref.~\cite{Casler:2019xqn} deals with a fourth-order polynomial equation that is solved using approximate functions.

\begin{table}[t!]
\begin{center}
\begin{tabular}{|c|c|c|c||c|}
\hline  
& Neutrino & Top quark \\
\hline
\hline This work & $(0.994, 0.991)$ & $(0.996, 0.998)$ \\
\hline Reference \cite{Casler:2019xqn} & $(0.967, 0.942)$ & $(0.974, 0.984)$ \\

\hline
\end{tabular}
\caption{Correlation between the simulated and reconstructed $(p_T, p_z)$ values in our analysis and ref.~\cite{Casler:2019xqn}.}
\label{tab:correlations}
\end{center}
\end{table}

\section{Method II: Improved event reconstruction}
\label{sec:MethodII}

In the previous section assuming that the $W^{\pm}$ bosons were on-shell and constrained their masses to a fixed value. By doing this, we obtained two extra equations, we were able to fully reconstruct the neutrino and antineutrino four-momenta. This assumption is motivated by the narrow width approximation, $\Gamma_{W} \rightarrow 0$ in the W-propagators
\begin{equation}
\frac{1}{ (P^2- m_{W}^2)^2  +  m_{W}^2 \Gamma_{W}^2 } \rightarrow \frac{\pi}{m_{W} \Gamma_{W}} \delta(P^2 - m_{W}^2) \, .
\label{eq:delta}
\end{equation}
While this approach has proven to be useful, its validity has certain limitations, as evidenced by the prediction of unphysical (imaginary) solutions for the neutrino four-momentum for a fraction ($ \sim 11 \%$) of the events. 

In this section, we develop a method to improve the reconstruction of the full event without relying on the assumption that the $W$ bosons have to be on-shell. Since our goal is to reconstruct the full $t \bar{t}$ events in the dileptonic channel, we also allow for the top quarks to be off-shell and analyse how this affects the reconstruction.

We remind the reader that we are reconstructing the events by using analytical expressions. Therefore, if we had truth-level information about the values of $(m_{W^+}, m_{W^-})$ then the reconstruction would be exact, up to  the combinatorial ambiguities and experimental errors. In the absence of additional information (e.g. polarization of initial and final states), the optimal $m_{W^{\pm}}$ values to reconstruct the event for a particular set of measured momenta $(P_{l}, P_{\bar l}, P_{j_1}, P_{j_2})$ are the ones that maximise the probability of measuring these values. Since this probability depends on the squared amplitude for the process, we suggest to use the following function in order to calculate the optimal $m_{W^{\pm}}$ values to perform the reconstruction
\begin{eqnarray}\label{eqn:K}
K(m_{W^{+}},  m_{W^{-}}) &=&  \frac{1}{ (m^{2}_{T}  - m_{t}^{2} )^{2} + (\Gamma_{t} m_{t})^{2}}   \; \;  \frac{1}{ (m^{2}_{\bar T}  - m_{t}^{2} )^{2} + (\Gamma_{t} m_{t})^{2}} \nonumber \\
&& \frac{N^{+}}{ (m^{2}_{W^{+}} - m_{W}^{2} )^{2} + (\Gamma_{W}  m_{W})^{2}}    \; \;   \frac{N^{-}}{ (m^{2}_{W^{-}} - m_{W}^{2} )^{2} +  (\Gamma_{W}  m_{W})^{2}}  \, ,
\end{eqnarray}
where $m_{T, \bar T}$ and $m_{W^{\pm}}$ are the reconstructed masses of the top quarks and $W$ bosons, respectively, and $m_{t}$ and $m_{W}$ denote their pole masses. The factors $N^{+}$ and $N^{-}$ are given by
\begin{equation}
N^{+} = (P_{\bar b} \cdot P_{\bar \nu })(P_{\bar t} \cdot P_{l}) \, , \quad N^{-} = (P_{b} \cdot P_{\nu})(P_{t} \cdot P_{\bar l}) \, ,
\end{equation}
and only appear when summing over the spins of the final-state leptons, $b$ and $\bar{b}$ quarks produced by the $t$ and $\bar{t}$ decays in the decay chain approximation\cite{Frixione:2007zp}. Note that besides the explicit depence on $(m_{W^{+}},  m_{W^{-}})$ in the function $K$, there is also an implicit dependence on these masses in $m_{T, \bar T} =  m_{T, \bar T} (m_{W^{+}},  m_{W^{-}})$, as $P_{\nu,  \bar \nu} =  P_{\nu,  \bar \nu}  (m_{W^{+}},  m_{W^{-}})$.

In the previous section we found that the reconstruction of the neutrino momentum is not unique. More concretely, there are four possible reconstructions due to the combinatorial ambiguities in the sign choice in eq. \eqref{eqn:Enu} and the two jet-assignments for the $b$ and $\bar{b}$ quarks. In the improved reconstruction method presented in this section, the optimal reconstruction is defined as the one that maximises the function $K(m_{W^{+}},  m_{W^{-}})$ introduced in eq. \eqref{eqn:K}.

It is worth mentioning that using top quark parameters (mass and width) to optimize the event reconstruction is consistent, since we are not aiming to evaluate these parameters but to reconstruct the kinematics of the full event. In fact, both the mass and the decay width will be measured with great accuracy in a future $e^+e^-$ collider \cite{Baer:2013cma, Charles:2018vfv, Abramowicz:2018rjq, Bambade:2019fyw, Fujii:2019zll} operating at the top quark pair-production threshold around $350 \; \GeV$. However, our technique could be applied to the study of observables related to the polarization of the top quark, the $W$ boson or the $Wtb$ vertex \cite{Aguilar-Saavedra:2017wpl, Aguilar-Saavedra:2018ggp, Durieux:2018tev}, which we will explore in detail in Section~\ref{sec:Applications}.
The approach followed here is inspired by the Matrix Element Method \cite{Kondo:1988yd,Kondo:1991dw,Dalitz:1991wa}, which was originally developed to determine dynamical parameters such as masses or decay widths from measured quantities taking uncertainties into account.  In this work we are interested in reconstructing the event kinematics rather than extracting these parameters, and we use the amplitude for the process to weight different $W^\pm$ invariant mass assumptions.

\begin{figure}[h!]
     \begin{center}
         \includegraphics[scale=0.522]{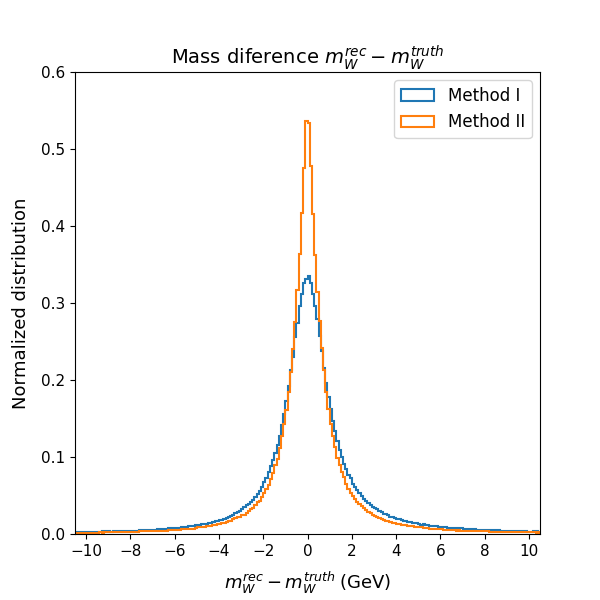}
         \includegraphics[scale=0.522]{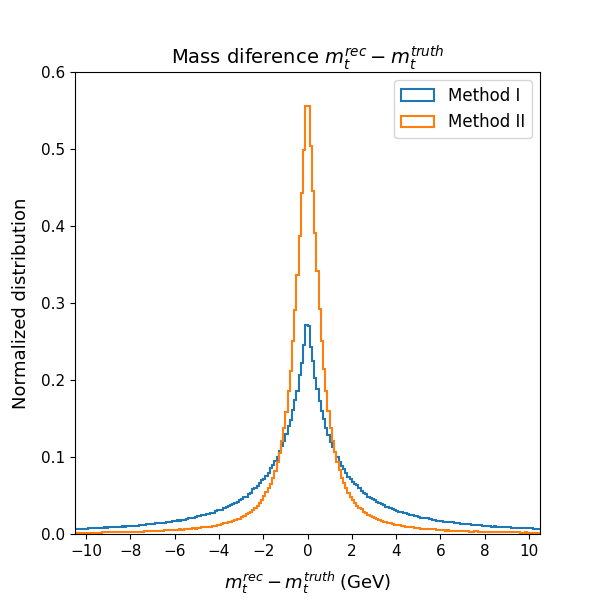}
      \end{center}
      \caption{On the left, we show the difference between the reconstructed and true $W^{\pm}$ masses, $m_{W}^{rec} - m_{W}$, for the two reconstruction methods. On the right, we do the same for the top quarks.}
    \label{fig:masses}
\end{figure}

The algorithm we used to implement the method presented in this section works as follows. First, we calculate the value of the function $K(m_{W^{+}},  m_{W^{-}})$ introduced in eq. \eqref{eqn:K} for the four possible reconstructions using $m_{W^{\pm}} = 80.4 \; \GeV$. If this value leads to an unphysical reconstruction in \eqref{eqn:Enu}, we search for the values of the masses that bring the discriminant to the positive region using a gradient descent algorithm with ADAM adaptative stepsize \cite{kingma2014adam}. Then we pick the solution with the smallest value as the correct physical reconstruction, and look for the absolute maximum of the function $K(m_{W^{+}},  m_{W^{-}})$ to find the optimal values of the masses for the reconstruction. If the second solution is less than $5$ times larger than the first solution\footnote{This parameter has been fixed to an optimal value.}, both are maximised and the one giving the largest value for $K$ is kept as the optimal reconstruction. Note that the function $K$ has at least $8$ maxima that correspond to the combinations of the poles in the denominators. Therefore, we use the differential evolution algorithm \cite{Storn1997}, as implemented in the \texttt{Optimize} package of the \texttt{SciPy} library \cite{scipy}, to search for the absolute maximum of $K$. Even though we are able to reconstruct  most of the $11 \%$ of the events with a negative discriminant, the quality of the reconstruction for these events is slightly reduced due to the maximization procedure. For this reason, we decided to obtain the most precise reconstruction and thus only present results for events with a positive discriminant. The code used to analyse the data can be found in this GitHub 
repository\footnote{\href{https://bit.ly/2wvxnVV}{https://github.com/pmramiro/tt$_-$reconstruction}
}.

\begin{figure}[h!]
   \begin{center}
        \includegraphics[scale=0.475]{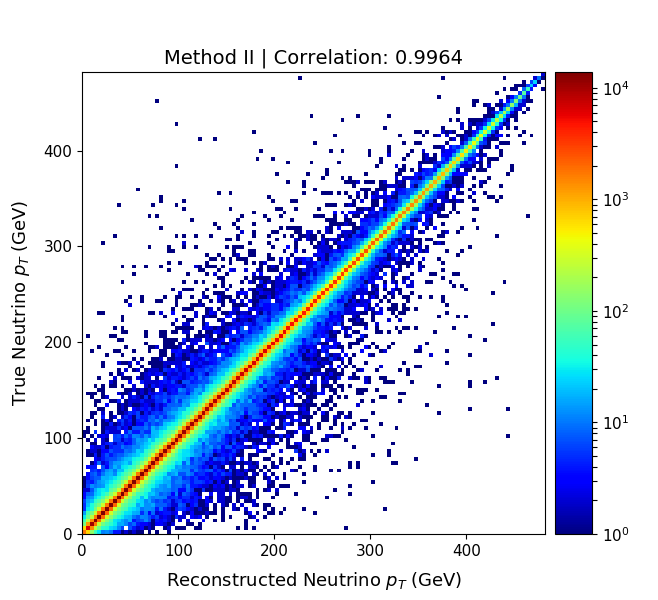}
        \includegraphics[scale=0.475]{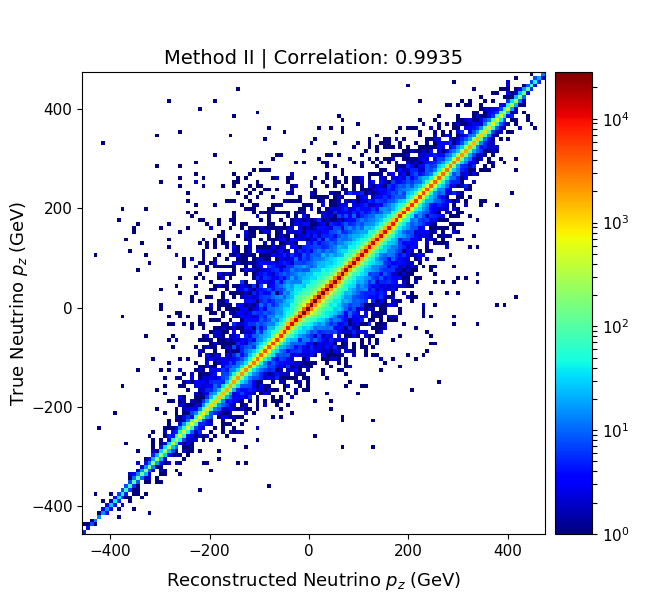}
        \includegraphics[scale=0.475]{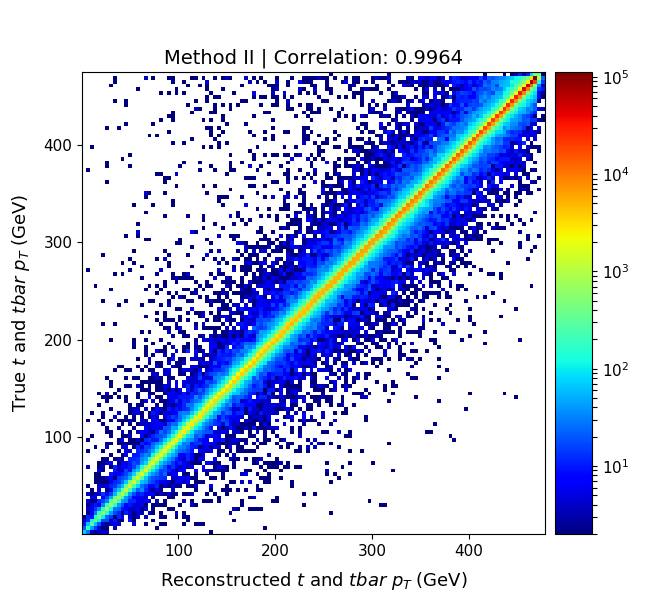}
        \includegraphics[scale=0.475]{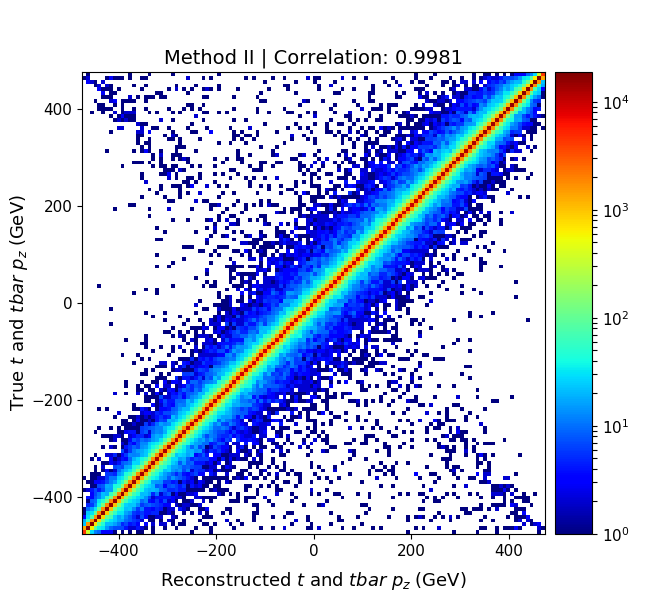}
     \end{center}
	\caption{Transverse (left) and longitudinal (right) components of the simulated neutrino (top panels) and top quarks (lower panels) spatial three-momenta as a function of the reconstructed momenta for a sample of $10^{6}$ events using Method II. The correlation between the simulated and reconstructed data is shown in every figure title.}
    \label{fig:p_improved}
\end{figure}

\begin{figure}[h!]
      \includegraphics[scale=0.55]{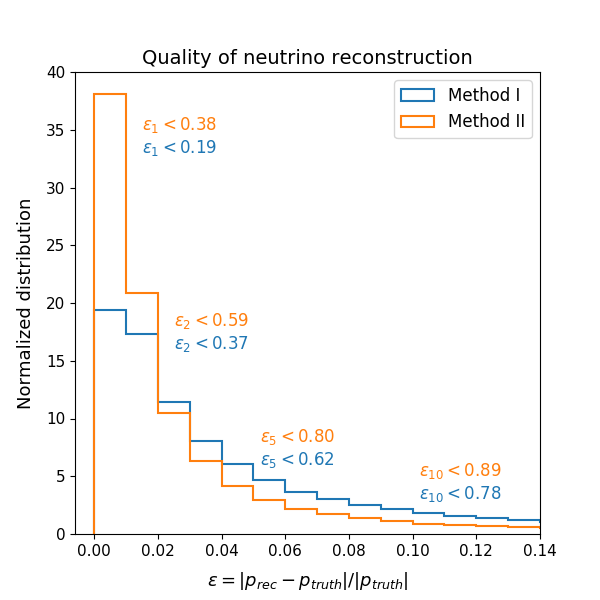}
      \includegraphics[scale=0.55]{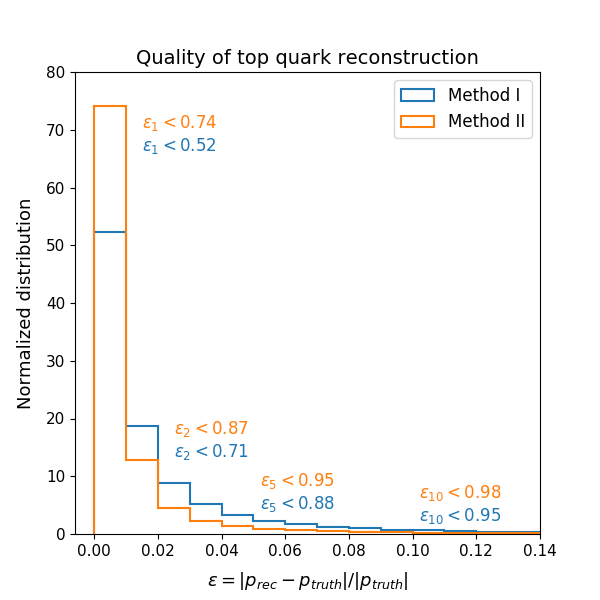}
	\caption{Quality of neutrino and top quark reconstructions for Method I and II. In each figure, $\epsilon_{x}$ denotes the fraction of events with a reconstruction error of less than $x \, \%$. Method II gives a substantial improvement in the reconstruction error compared to Method I.}
    \label{fig:quality}
\end{figure}

We now present our results for the reconstruction of the events using the improved Method II. First, we show the difference between the reconstructed and true masses for the $W$ bosons and top quarks in Figure~\ref{fig:masses}. Interestingly, we can observe that Method II significantly outperforms Method I in the reconstruction of the $W$ boson masses and, as a consequence, in the top quark masses. In Figure~\ref{fig:p_improved} we compare again the reconstructed neutrino and top quark momenta with the truth-level information from the simulated events. It is apparent that there is a slight improvement with respect to Method I, as confirmed by the correlation coefficient given at the top of each figure. However, since the relative error is small in both methods  this variable is not the best to compare the quality of the two methods. For this reason, we introduce a different performance metric in Fig. \ref{fig:quality}, the error  $\epsilon = |\vec{p}_{\text{rec}} - \vec{p}_{\text{truth}}| / |\vec{p}_{\text{truth}}|$. Using this performance metric we see again that Method II considerably outperforms Method I, and we show the performance of both methods under this quantity in Table ~\ref{tab:performance}. In particular, while both methods perform well for $\epsilon < 10 \%$, specially for the top quark reconstruction, Method II performs much better for $\epsilon < 5 \%$ and provides an excellent reconstruction for top quarks in $87 \%$ of the events, with an error of less than $2 \%$.

\begin{table}[t!]
\begin{center}
\begin{tabular}{|c|c|c|c|c|c|}
    
\hline
 & Reconstruction & $\epsilon < 1 \%$ & $\epsilon < 2 \%$ & $\epsilon < 5 \%$ &  $\epsilon < 10 \%$ \\
\hline
\multirow{2}*{Neutrino} & Method I & $19 \%$ & $37 \%$ & $62 \%$ & $78 \%$ \\
                        & Method II & $38 \%$ & $59 \%$ & $80 \%$ & $89 \%$ \\
\hline
\hline
\multirow{2}*{Top quark} & Method I & $52 \%$ & $71 \%$ & $88 \%$ & $95 \%$ \\
                         & Method II & $74 \%$ & $87 \%$ & $95 \%$ & $98 \%$ \\
\hline

\end{tabular}
\caption{Reconstruction error $\epsilon$ for Method I and II for neutrinos and top quarks. Columns third to sixth show the percentage of events reconstructed with and error $\epsilon < 1, 2, 5, 10 \, \%$, respectively. }
\label{tab:performance}
\end{center}
\end{table}

\newpage
\mbox{~}
\newpage

\section{Application: Helicity fractions of the  W bosons from top quark decays }

\label{sec:Applications}

In the previous sections we developed two methods to perform an analytical reconstruction of the $t\bar{t}$ pair in the dilepton channel. We  demonstrated that both techniques can be used to achieve an accurate reconstruction. In this section, we apply these reconstruction methods to extract the decay rates of (unpolarized) top quarks into longitudinally and transversally polarized $W$ bosons. A precise measurement of these observables could be used to probe new physics effects in the $Wtb$ vertex.

The helicity fractions of the $W$ boson can be studied through the analysis of the angular distribution of the top quark decay products in the leptonic channel.
Let  $\theta^{*}$  be the angle between the direction of the charged lepton arising from the $W$ boson decay and the reversed direction of the the top quark, both in the rest frame of the $W$ boson. The normalized differential decay rate for unpolarised top quarks can be written as
\begin{equation}
\frac{1}{\Gamma} \frac{d\Gamma}{d\cos\theta^{*}} = \frac{3}{4} \left(1-\cos^{2}\theta^{*} \right)F_{0} + \frac{3}{8} \left(1-\cos\theta^{*} \right)^{2}F_{L} + \frac{3}{8} \left(1+\cos\theta^{*} \right)^{2}F_{R} \, ,
\label{eq:theta}
\end{equation}
where $F_i$ denote the helicity fractions, and $F_{0}+F_{L}+F_{R} = 1$. At  leading order, the SM prediction in the $m_b=0$ limit only depends on $m_W/m_t$. In particular, $F_{R} =0 $ vanishes due to the $V-A$ structure of the $Wtb$ vertex. However, there are many new physics models that could modify this vertex, inducing anomalous top couplings.

\begin{figure}[h!]
   \begin{center}
     \includegraphics[scale=0.68]{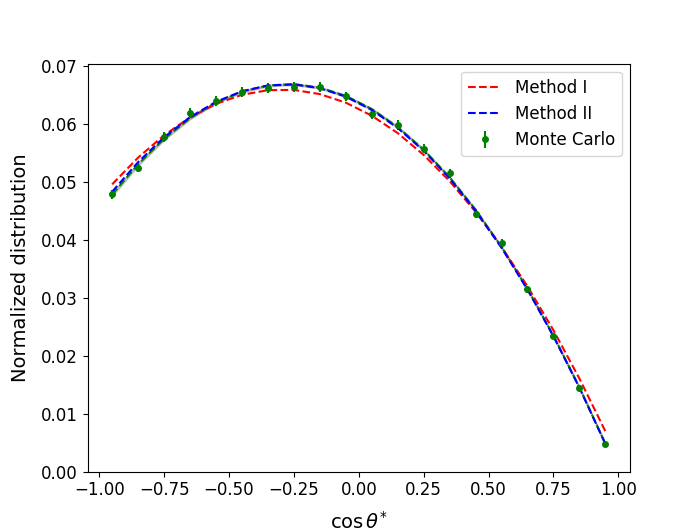}
   \end{center}
\caption{Distribution of $\cos\theta^{*}$  from Monte Carlo data and from Method I  \&  Method II reconstructions.}
    \label{fig:helicity}
\end{figure}

In Figure ~\ref{fig:helicity}, we show results for the angular distribution presented in eq. \eqref{eq:theta} using the two reconstruction methods developed in this work. After reconstructing all the events, we binned data  and calculated the helicity fractions by doing a 3-parameter fit to the distribution from eq. \eqref{eq:theta}. The statistical error was estimated assuming that the data are sampled from a Gaussian distribution. We checked that taking different bin sizes does not significantly affect the fit result. From Fig. ~\ref{fig:p_improved},  it is clear that both Method I and II provide a precise prediction for the helicity fractions, as was expected given the high accuracy of the neutrino four-momentum reconstruction.

Detailed results are given in Table~\ref{tab:helicity}. In particular, the helicity fractions computed using reconstruction Method II are consistent with the calculation using the truth-level information from the Monte Carlo simulation. For comparison, we also show the helicity fractions from the tree level SM prediction using the $(m_t, m_W, m_b)$ input values from the Monte Carlo data. In the narrow width approximation, the helicity fractions extracted from the LO Monte Carlo simulation should approach the tree level SM prediction, as reflected in Table ~\ref{tab:helicity}. For completeness, we also include the combined experimental results obtained by ATLAS and CMS \cite{Aaboud:2016hsq, Khachatryan:2016fky} and the NNLO SM prediction \cite{ PhysRevD.45.124, Czarnecki:2010gb}.

\begin{table}[t!]
\begin{center}
\begin{tabular}{|c|c|c|c|}
\hline
& $F_{0}$ & $F_{L}$ & $F_{R}$ \\
\hline 
\multicolumn{4}{c}{}\\[-4mm]
\hline
Monte Carlo data & $0.702 \pm 0.001$ & $0.299 \pm 0.001$ & $-0.002 \pm 0.0003$ \\
\hline
Method I & $0.672 \pm 0.002$ & $0.312 \pm 0.001$ & $\hspace{6.5pt} 0.015 \pm 0.001$ \\
\hline
Method II & $0.699 \pm 0.001$ & $0.302 \pm 0.001$ & $-0.002 \pm 0.0003$ \\
\hline 
\multicolumn{4}{c}{}\\[-4mm]
\hline
SM LO & 0.698 \phantom{xxxxxx} & 0.302   \phantom{xxxxxx} & 0.000  \phantom{xxxxx} \\
\hline 
\multicolumn{4}{c}{}\\[-4mm]
\hline
SM  NNLO QCD & $0.687 \pm 0.005$ & $0.311 \pm 0.005$ & \hspace{6.5pt}  $ 0.0017 \pm 0.0001$  \\
\hline
ATLAS \& CMS & $0.695 \pm 0.032$ & $0.311 \pm 0.022$ & $-0.006 \pm 0.020$ \\
\hline
\end{tabular}
\caption{Helicity fractions calculated using truth level information (i.e. Monte Carlo simulation), Method I and II, SM prediction and the combined ATLAS and CMS results.}
\label{tab:helicity}
\end{center}
\end{table}

Finally, it is worth mentioning that we do not expect large systematic errors coming from the reconstruction at partonic level. However, one might expect sizeable systematic uncertainties once detector effects are included, whose impact will be studied in detail in future works.

\newpage
\section{Conclusions}
\label{sec:conclusions}

One of the main goals of a future $e^+e^-$ collider is the study of top quark properties \cite{Abramowicz:2018rjq, Fujii:2019zll}. The physics program of both CLIC and ILC includes precision measurements of the top quark mass, its electroweak and Yukawa couplings, anomalous couplings, polarization, and its potential interactions with new physics particles. In order to perform such measurements, the dilepton channel of the $t \bar{t}$ events offers a clean final state, but there are two potential challenges: performing the kinematic reconstruction of the full event in the presence of two neutrinos and the ambiguity in the $b$ and $\bar{b}$ jets assignment.

In this work, we have developed two methods for the analytical reconstruction of the $t \bar{t}$ pair in the dilepton channel. Since the only unknowns are the three-momenta of the two neutrinos and energy-momentum conservation in the event provides four equations, two more inputs are needed for the reconstruction. In the first approach (Method I), we assumed that the $W^{\pm}$ bosons mediating the decay are on-shell and their masses are fixed to the pole value, and derived a simple and compact analytical expression to perform the reconstruction. As a consequence, the reconstruction algorithm is fast and easy to implement. This method provides a highly accurate reconstruction of both the neutrino and top quark four-momenta, with relative errors of less than $5 \%$ for $62 \%$ and $80 \%$ of the events, respectively.

In the second approach (Method II), we have improved our results calculating the optimal values of the $W^{\pm}$ masses to perform the reconstruction. That is, we took the $m_{W^\pm}$ values that maximise the probability of measuring the observed set of four-momenta: $(P_l, P_{\bar l}, P_{j_1}, P_{j_2})$. The results obtained using Method II are extremely precise, at the cost of longer computational times. In particular, we were able to reconstruct the neutrino and top quark four-momenta with a relative error of less than $5 \%$ for $80 \%$ and $95 \%$ of the events, respectively, and with an error of less than $1 \%$ for $38 \%$ and $74 \%$ of the events.

Finally, we have studied how this technique could be applied to study the polarization of the $W$ boson and the $Wtb$ vertex. Using our reconstruction method, the helicity fractions of the $W$ boson can be calculated with a precision of less than $1 \%$. 
Compared with previous studies in the literature, we conclude that the algorithms described here are not only fast and easy to implement, but also lead to a better performance.

\section{Acknowledgements}

We are grateful to J. A. Aguilar-Saavedra for helpful discussions and suggestions, and to J.~R. Espinosa for critically reading the manuscript.  We would like to acknowledge support from the Spanish Research Agency (Agencia Estatal de Investigaci\'on) through the contract FPA2016-78022-P and IFT Centro de Excelencia Severo Ochoa under grant SEV-2016-0597.

\bibliographystyle{j-cerdeno}
\bibliography{references}

\end{document}